\documentclass[12pt]{article}
\usepackage{graphicx}

\def\ignore#1{{}}

\let\oldtheequation=\theequation
\def\doteqs#1{\setcounter{equation}{0}            
\def\theequation{{#1}.\oldtheequation}}
\newcounter{sxn}
\def\sx#1{\addtocounter{sxn}{1} \vskip 1.cm  \goodbreak
\noindent{\large\bf\leftline{\thesxn.~~#1}} \nobreak \vskip -.5cm}
\def\sxn#1{\sx{#1} \doteqs{\thesxn}}

\newcounter{axn}

\date{}

\newdimen\mybaselineskip
\newcommand{\beeq}{\begin{equation}}
\newcommand{\eneq}{\end{equation}}
\newcommand{\beqn}{\begin{eqnarray}}
\newcommand{\eeqn}{\end{eqnarray}}

\def\la{\raise.16ex\hbox{$\langle$}\lower.16ex\hbox{}  }
\def\ra{\, \raise.16ex\hbox{$\rangle$}\lower.16ex\hbox{} }

\def\psibar{ \psi \kern-.65em\raise.6em\hbox{$-$} \lower.6em\hbox{} }
\def\psibarb{ \psi \kern-.65em\raise.6em\hbox{$-$}  }

\begin{document}

\thispagestyle{empty}

\baselineskip=12pt

%{\small \noindent \mydate  \hfill UW}

%{\small \noindent Ramin\hfill   }

\vspace*{3.cm}

\begin{center}  
{\LARGE \bf  High Overtone Quasinormal Modes of Analog Black Holes and the Small Scale Structure of the Background Fluid}
\end{center}

\baselineskip=14pt

\vspace{3cm}
\begin{center}
{\bf  Ramin G. Daghigh$\sharp$ and Michael Green$\dagger$}
\end{center}

\centerline{\small \it $\sharp$ Natural Sciences Department, Metropolitan State University, Saint Paul, Minnesota, USA 55106}
\vskip 0 cm
\centerline{} 

\centerline{\small \it $\dagger$ Mathematics Department, Metropolitan State University, Saint Paul, Minnesota, USA 55106}
\vskip 0 cm
\centerline{} 

\vspace{1cm}
\begin{abstract}
The goal of this paper is to build a foundation for, and explore the possibility of, using high overtone quasinormal modes of analog black holes to probe the small scale (microscopic) structure of a background fluid in which an analog black hole is formed.  This may provide a tool to study the small scale structure of some interesting quantum systems such as Bose-Einstein condensates.
%More specifically, one may be able to use these modes as a tool to explore the microscopic structure of interesting quantum fluids such as Bose-Einstein condensate studied in this paper. 
In order to build this foundation, we first look into the hydrodynamic case where we calculate the high overtone quasinormal mode frequencies of a $3+1$ dimensional canonical non-rotating acoustic black hole.  The leading order calculations have been done earlier in the literature.  Here, we obtain the first order correction. We then analyze the high overtone quasinormal modes of acoustic black holes in a Bose-Einstein condensate using the linearized Gross-Pitaevskii equation.   We point out that at the high overtone quasinormal mode limit, the only term that is important in the linearized Gross-Pitaevskii equation is the quantum potential term, which is a small scale effect. 
\baselineskip=20pt plus 1pt minus 1pt
\end{abstract}

%%%%%%%%%%% 1 %%%%%%%%%%
\newpage

\sxn{Introduction}

\vskip 0.3cm

Quasinormal modes (QNMs) of black holes are the damped vibrational modes of the perturbed black hole spacetime exterior to the event horizon.  
%The frequency spectrum is discrete and complex.  The imaginary part of the frequency indicates the presence of damping, which is a consequence of the boundary conditions that require energy to be carried away from the system.  
Throughout this paper, we use the symbol $\omega$ to indicate the complex QNM frequency and the symbols $\omega_R$ and $\omega_I$ are used to indicate the real part and imaginary part (damping rate) of this frequency respectively.

\begin{figure}[tb]
\begin{center}
\includegraphics[height=8cm]{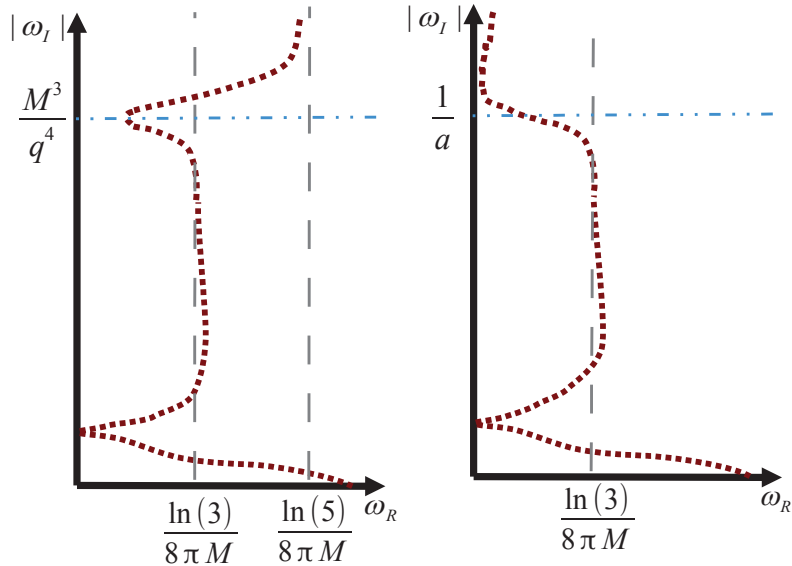}
\end{center}
\caption{ A rough schematic behavior of the QNM frequency spectrum for a Reissner-Nordstr$\ddot{\rm o}$m black hole (left) with small charge, $q$, and Kerr black hole (right) with small angular momentum per unit mass, $a$.  The spectrum deviates significantly from the Schwarzschild spectrum only at large damping rates where $|\omega_I| > M^3/q^4$ for the Reissner-Nordstr$\ddot{\rm o}$m case and $|\omega_I| > 1/a$ for the Kerr case.  $M$ is the mass of the black hole.}
\label{QNM-spec-Kerr-RN}
\end{figure}

\begin{figure}[tb]
\begin{center}
\includegraphics[height=8.5cm]{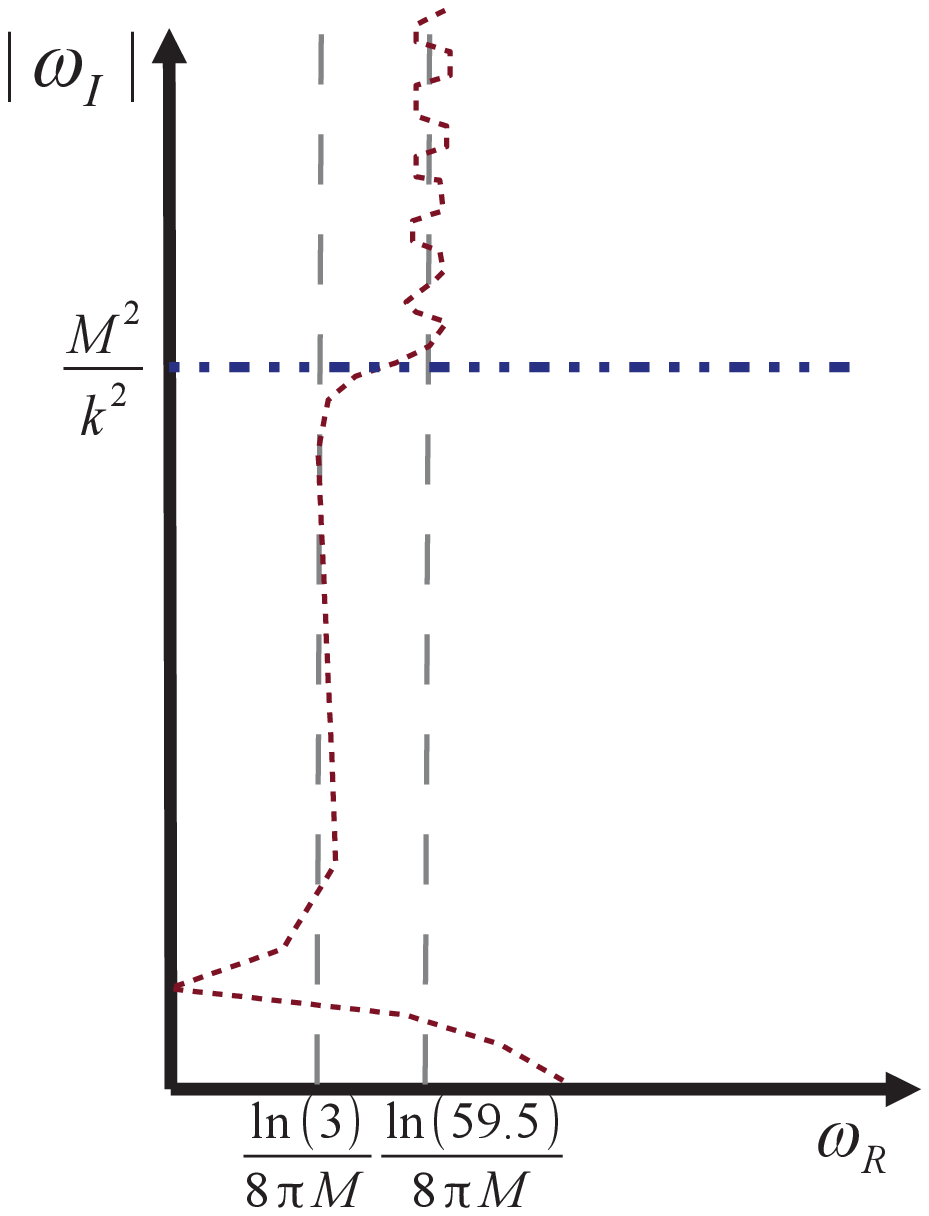}
\end{center}
\caption{ A rough schematic behavior of QNM frequency spectrum for the quantum-corrected black hole model suggested in \cite{Peltola-K}.  The spectrum deviates significantly from the Schwarzschild spectrum only at large damping rates where $|\omega_I| > M^2/k^2$.  $M$ is the mass of the black hole and $k$ is the polymerization (Planck) length scale. }
\label{QNM-spec-PK-BH}
\end{figure}

Black hole QNMs with low damping rates are important for gravitational wave observations since they describe the frequency spectrum of the gravitational radiation that is expected to emerge from black hole formation during late times.  The high overtone QNMs,
while not in practice observable due to their short relaxation time, became a subject of interest
due in part to a conjecture by Hod\cite{Hod} relating them to the spacing between semiclassical
area/entropy eigenvalues of the quantum black hole. 
The possibility of a connection between the high overtone QNMs of black holes and quantum gravity remained conjectural until a link was established in \cite{BDK} between
the high overtone QNMs and the small scale structure of black hole spacetimes. The work in \cite{Andersson} and \cite{Ramin-RN} on the highly damped QNMs of Reissner-Nordstr$\ddot{\rm o}$m black holes in the small charge limit and the work done in \cite{Ramin-Kerr} on the highly damped QNMs of Kerr black holes in the small angular momentum limit show that the high overtone QNMs are sensitive
to any small scale (such as a small charge or angular momentum) that is added to
a Schwarzschild metric.  This realization led the authors in \cite{BDK} to conclude that any additional length scale due to a quantum correction should also modify the QNMs specifically in the high damping regime.
This was demonstrated explicitly in \cite{BDK} by calculating the highly damped
QNMs of the quantum corrected Schwarzschild black hole derived in \cite{Peltola-K}.  As shown schematically in Figure \ref{QNM-spec-Kerr-RN}, adding a small charge, $q$, or a small angular momentum per unit mass, $a$, to a Schwarzschild black hole of mass $M$ can only have a significant impact on the QNM frequency spectrum when $|\omega_I| > M^3/q^4$ for the Reissner-Nordstr$\ddot{\rm o}$m case and $|\omega_I| > 1/a$ for the Kerr case.  It was shown explicitly in \cite{BDK} that adding a small quantum correction, which introduces a small length scale of the order of Planck length, will modify the QNM spectrum only when  $|\omega_I| > M^2/k^2$, where $k$ is the polymerization (Planck) length scale.  This is shown schematically in Figure \ref{QNM-spec-PK-BH}.

The connection between high overtone QNMs and the small scale structure of black holes seems to be analogous to the toy model of a stretched string fixed at two ends.  Such a system, when damping is ignored, has a discrete resonance frequency spectrum called the normal modes.  The modes with low frequency, which have a wavelength comparable to the length of the string, will be unaffected by the small scale molecular structure of the string, but the high frequency normal modes with wavelengths comparable to the spacing between the molecules will be affected significantly by the small scale molecular structure of the string.

The established connection between the high overtone QNMs and the small scale structure of black holes in \cite{BDK} motivates us to investigate if there exists a similar connection between the high overtone QNMs of {\it analog} black holes and the small scale structure of some interesting quantum systems, such as Bose-Einstein condensates (BECs).

The first analog black hole was experimentally realized in a BEC in $2009$\cite{Lahav}.  Analog black holes are used as the experimental field for Hawking radiation.   Astrophysical black holes obey the uniqueness theorem for black hole spacetimes and they do not have ``hair'' other than mass, angular momentum and charge.  Analog black holes on the other hand have numerous hairs due to their geometric correspondences.  Geometries of analog black holes are determined by factors like the equations of state, boundary conditions on fluid systems, etc.  Since the Hawking radiation in analog black holes is influenced by the background geometry or wave equation on the analog curved spacetime, it becomes important that one determines if the required geometry is present or not.  This can be done through the detection of QNMs, which are determined by geometric factors of the analog black holes.  This is the main motivation for studying QNMs in the context of analog black holes.  

By investigating the high overtone QNMs of analog black holes in BECs, we want to determine the possibility and practicality of using these modes as a viable experimental tool to gain information about the small scale structure of these interesting quantum systems.  This will enable us to understand and formulate the macroscopic behavior of these many-body systems in terms of the behavior of their individual components.
%One condition for the practicality of using high overtone QNMs in this context is that the real part of the frequency should approach a finite value, similar to the Schwarzschild case, so that one can experimentally produce them in the quantum fluid under consideration.  
A method to investigate the small scale structure (the density distribution of the atoms) of BEC is developed by Gericke et al.\ in \cite{Gericke} using electron impact ionization.  In this method electrons are used to detect the atoms in the BEC by ionizing them.  In this paper, we take another approach using QNMs.
If it is possible to produce these modes in an experimental setting, it will become possible to test theoretical models of BECs in the small scale limit.
%be interesting to see how different microscopic features of the BEC can affect the Gross-Pitaevskii equation and consequently the high overtone QNMs.  It may also be possible to discover new small scale effects, which has not been theorized yet. As an example, an intriguing microscopic correction to the Gross-Pitaevskii equation can be due to the Generalized Uncertainty Principle (see for example \cite{Maggiore-GUP} for one of the early papers on this topic), which is a quantum gravity correction to Heisenberg's Uncertainty Principle.  
Not many calculations can be found in the literature on the QNMs of analog black holes in the background of a BEC.  The first of these calculations\cite{Nakano} is done for a BEC with a one-dimensional background velocity and a density with a power or a combination of powers of the spatial coordinate along which the background velocity is directed.  The authors of \cite {Nakano} use the linearized Gross-Pitaevskii equation in the low frequency limit, where the quantum potential can be ignored, in which case one is left with the hydrodynamic dispersion relation.   This approximation is not valid for high overtone QNMs and for that reason is not relevant to this paper.  In addition, Barcelo et al.\cite{Barcelo, Barcelo1} calculate the QNMs of an analog black hole in the background of a BEC with one-dimensional flow and step-like discontinuities at the sound horizon.  They keep the quantum potential in  their linearized Gross-Pitaevskii equation and consider the full Bogoliubov dispersion relation (not the hydrodynamic approximation as in \cite{Nakano}).  Therefore, in this paper we focus on the work of \cite{Barcelo, Barcelo1}.

Our main motivation here is to explore a connection between high overtone QNMs of analog black holes and the small scale structure of the background medium in which they are formed. In order to do this we need to be able to compare our results, in the appropriate limit, to the hydrodynamic case.  For that reason, we improve upon the known results in the literature for high overtone QNMs of analog black holes in the background of a continuous fluid (hydrodynamic limit) where the small scale structure becomes irrelevant.  In the high overtone limit, not much has been done in the literature since there has not been a strong motivation for this limit.  Berti et al.\cite{Berti-sonic} calculated the high overtone QNMs for a rotating acoustic $2+1$ dimensional black hole (the ``draining bathtub" case) and they showed that the high overtone limit does not exist for this case.  In \cite{Saavedra}, the QNM spectrum of
Unruh's $3 + 1$ dimensional acoustic black hole has been found to be purely imaginary, which makes such a black hole unstable under sonic perturbations.  On the other hand, it was shown in \cite{Berti-sonic} that the high overtone QNMs of a canonical non-rotating $3+1$ dimensional black hole, which is a stable configuration, does exist and in this large damping limit the real part of the frequency approaches a finite value.  Therefore, this case would be the most interesting case in the context of an experimental setting, since it would be plausible to create these modes in a lab and explore their connection to the microscopic structure of a known fluid.  Therefore, in this paper we will be focusing on the high overtone QNMs of a $3+1$ dimensional canonical non-rotating  black hole.

The outline of the paper is as follows. In Section 2, we look into the hydrodynamic case and calculate the leading and first order high overtone QNM frequency for a $3+1$ dimensional canonical non-rotating acoustic black hole formed in a fluid background.  In section 3, we analyze the high overtone QNMs of a $1+1$ dimensional analog black hole in the background of a BEC. In Section 3, we present our summary and conclusions.

\sxn{High Overtone QNMs: The Hydrodynamic Case}

The leading order high overtone QNMs of a $3+1$ dimensional canonical non-rotating acoustic black hole formed in a fluid are calculated in \cite{Berti-sonic}.  In this section, we calculate the first order correction of the high overtone QNMs for a $3+1$ dimensional canonical non-rotating acoustic black hole using the method developed in \cite{M-S}.  The QNM wave equation for a $3+1$ dimensional acoustic black hole is\cite{Berti-sonic}

\beeq
\partial_{r_*}^2{\Phi}+\left[ {\omega^2 \over c^2}-V(r) \right]{\Phi}=0 ~
\label{schrodinger}
\eneq
where  
\beeq
V(r)= \left(1-{r_0^4 \over r^4}\right)\left[{l(l+1) \over r^2}+{4 r_0^4 \over r^6}\right]~
\label{potential}
\eneq
$c$ is the speed of sound, $r$ is the radial coordinate, $r_0$ is a normalization constant and $l$ is the angular number.  The coordinates $r_*$ and $r$ are related according to   
\beeq
{dr_* \over dr}={1 \over f(r)}~
\label{r*-r}
\eneq
where
\beeq
f(r)=1-{r_0^4 \over r^4}~.
\eneq
Note that the potential $V(r)$ approaches zero when $r\rightarrow r_0$ ($r_* \rightarrow -\infty$) and $r\rightarrow \infty$ ($r_* \rightarrow \infty$).  Therefore, the solutions at the boundaries take the form $e^{\pm i{\omega \over c} r_*}$.  If we assume the perturbations depend on time as $e^{i\omega t}$ ($\omega_I >0$), then the outgoing boundary conditions for damped modes (QNMs) are
\begin{eqnarray}
&& e^{ i{\omega \over c} r_*},~~r\rightarrow r_0~(r_* \rightarrow -\infty)   \nonumber \\
&& e^{-i{\omega \over c} r_*},~~r\rightarrow \infty~(r_* \rightarrow \infty)~.  
\end{eqnarray}
In the high damping limit, the wave equation (\ref{schrodinger}) can be approximated to leading order as
\beeq
\partial_{r_*}^2{\Phi}+\left[ {\omega^2 \over c^2}-{(j^2-1) \over 4r_*^{2}}\right]{\Phi}=0 ~
\eneq
where $j=3/5$. 

\begin{figure}[tb]
\begin{center}
\includegraphics[height=8cm]{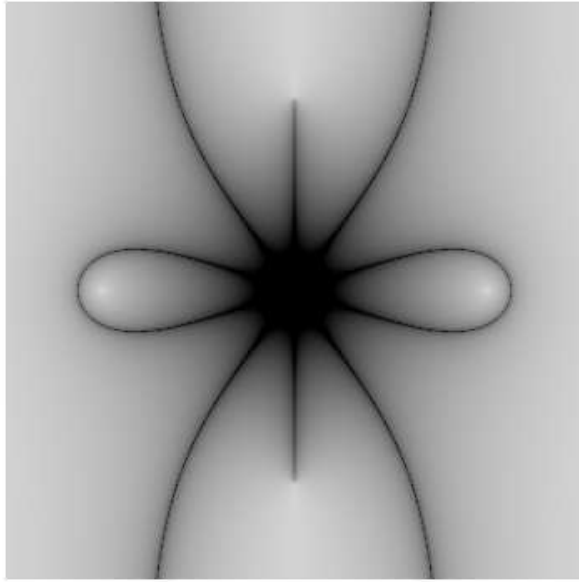}
\end{center}
\caption{The structure of the anti-Stokes lines in the infinite damping limit for a $3+1$ dimensional canonical non-rotating acoustic black hole.}
\label{3+1-BH}
\end{figure}

%In order to investigate the structure of anti-Stokes lines in the $r$-plane, we rescale the wavefunction as
Our calculations will rely on the structure of anti-Stokes lines in the complex $r$-plane.  To determine this structure, we rescale the wavefunction as
\beeq
\phi=\sqrt{f} \Phi~.
\eneq
This leads to the wave equation
\beeq
\partial_{r}^2{\phi}+R(r){\phi}=0 ~
\eneq
where
\beeq
R(r)=\left({1 \over f}\right)^2\left[ {\omega^2 \over c^2}-V(r)+{1\over 4} (\partial_{r}f)^2-{1\over 2}f \partial_{r}^2f \right] ~.
\eneq
The WKB solutions to this wave equation are
\beeq
\phi_{1,2}= R^{-1/4}\exp \left[ \pm i \int_t^r \sqrt{R(r')}dr' \right] \rightarrow \left({\omega \over cf}\right)^{-1/2}\exp \left[ \pm i {\omega \over c}\int_0^r {dr' \over f(r')} \right]~
\eneq
in the high damping limit ($|\omega| \rightarrow \infty$) as long as one stays away from the pole at the origin ($r=0$).  (The lower limit of integration $t$ is taken to be one of the zeros of $R$.) Therefore, in this limit, the anti-Stokes lines are the lines in the complex $r$-plane on which $ {\omega \over c}\int_0^r {dr' \over f(r')}(= {\omega \over c} r_*)$ is purely real.
%In the infinite damping limit, this wave equation can be approximated as
%\beeq
%\partial_{r}^2{\phi}+\left({1 \over f}\right)^2\left[ {\omega^2 \over c^2}-{(25j^2-1) r_0^8\over 4r^{10}}\right]{\phi}=0 ~,
%\label{}
%\eneq
%where $j=3/5$. 
The structure of the anti-Stokes lines, which originate from and/or end at $r=0$, in the infinite damping limit is shown in Fig.\ \ref{3+1-BH}.

In the large damping limit, the potential $V[r(r_*)]$ is negligible everywhere in the complex plane except in the vicinity of $r=r_*=0$.  To explore the behavior of the potential near $r=r_*=0$, we find the Taylor expansion of the right hand side of Eq.\ (\ref{r*-r}) around $r=0$ and then integrate, which gives
\beeq
r_*= -r_0 \sum_{k=0}^\infty {(r/r_0)^{4k+5} \over 4k+5}~.
\label{sumform}
\eneq  
After introducing a new variable
\beeq
z={\omega \over c} r_*~
\eneq 
we use Lagrange inversion of (\ref{sumform}) to find $r$ in terms of $z$:
\beeq
r= \left(-5r_0^4 {z \over \omega/c}\right)^{1\over 5 } \left\{1+{1\over -9}\left(-{5z\over r_0\omega/c}\right)^{4 \over 5 } +{10\over 1053}\left(-{5z\over r_0\omega/c}\right)^{8 \over 5 } +\cdots \right\}~.
\eneq 
Plugging $r(z)$ back into the potential given in Eq.\ (\ref{potential}) leads to
\beeq
V(z)= -{(\omega/c)^2 \over 4z^2} \left\{1-j^2+{25 (1-j^2) +36l(l+1)\over 225}\left({{5z\over -r_0\omega/c}}\right)^{4\over 5} + \cdots\right\}~.
%\label{}
\eneq 
We now look for solutions to the wave equation
\beeq
\partial_{z}^2{\Phi}+\left[ 1-{V(z) \over (\omega/c)^2} \right]{\Phi}=0 ~.
\label{waveeq-z}
\eneq
To go beyond leading order, we expand the wavefunction in $1 \over (-r_0\omega/c)^{4/5}$,
\beeq
\Phi = \Phi^{(0)}+{1 \over (-r_0\omega/c)^{4/5}}\Phi^{(1)} +{\rm O}\left(   \omega^{-{8 \over 5}}  \right)~.
%\label{}
\eneq
Using the expansion above, the zeroth (leading) order wave equation is found to be
\beeq
\partial_{z}^2{\Phi^{(0)}}-\left[ 1-{j^2-1 \over 4z^2} \right]{\Phi^{(0)}}=0 ~
%\label{}
\eneq
and the first order wave equation is
\beeq
\partial_{z}^2{\Phi^{(1)}}+\left[ 1-{j^2-1 \over 4z^2} \right]{\Phi^{(1)}}=(-r_0\omega/c)^{4 \over 5} \delta V \Phi^{(0)} ~
%\label{}
\eneq
where 
\beeq
\delta V=-{j^2-1 \over 4z^2}+{1\over (\omega/c)^2}V(z) ~.
%\label{}
\eneq
Solutions to the zeroth order wave equation are
\beeq
\Phi_\pm^{(0)}=\sqrt{\pi z \over 2}J_{\pm{j\over 2}}(z) ~.
\label{0th-order-solution}
\eneq
To have the correct behavior at the boundary where $z\rightarrow \infty$ along the positive real axis in the $z$-plane, we need a linear combination of the above solutions of the form
\beeq
\Phi^{(0)}= \Phi_+^{(0)}- e^{-i\pi j/2} \Phi_-^{(0)} \mathop{\sim}_{z\rightarrow \infty} -e^{-i\pi (1+j)/4} \sin(\pi j/2) e^{-iz}~.
%\label{}
\eneq
The solution to the first order wave equation is
\beeq
\Phi_\pm^{(1)}=\mathcal{C} \Phi_+^{(0)}(z)\int_0^zd\eta \Phi_+^{(0)}\delta V \Phi_\pm^{(0)}   - \mathcal{C}  \Phi_-^{(0)}(z)\int_0^zd\eta \Phi_+^{(0)}\delta V \Phi_\pm^{(0)}~
%\label{}
\eneq
where
\beeq
\mathcal{C} ={(-r_0\omega/c)^{4 \over 5} \over \sin(\pi j/2)} ~
%\label{}
\eneq
and the integrations are done along the positive real axis on the $z$-plane.
Therefore, the perturbative general solution to the wave equation (\ref{waveeq-z}) can be written as 
\beeq
\Phi=A\left( \Phi_+^{(0)}+{1\over (-r_0\omega/c)^{4\over 5}}\Phi_+^{(1)}+\cdots \right)+   B\left( \Phi_-^{(0)}+{1\over (-r_0\omega/c)^{4\over 5}}\Phi_-^{(1)}+\cdots \right)~.
\label{Phi-original}
\eneq
To make it easier to apply the boundary condition at infinity, we want to write this in the form $\Phi=\Phi^{(0)}+\omega^{-4/5} (\ldots)$. If 
we choose $A=1$ and $B=-\left(1-{\xi \over (-r_0\omega/c)^{4\over 5}}\right) e^{-\pi i j/2}$, we get
\beeq
\Phi=\Phi^{(0)}+  {1\over (-r_0\omega/c)^{4\over 5}} \left( \Phi_+^{(1)}-e^{-\pi i j/2}\Phi_-^{(1)}+\xi e^{-\pi i j/2} \Phi_-^{(0)}\right)+\cdots~.
%\label{}
\eneq
Once again, we require that this function satisfy the boundary condition $\Phi \rightarrow e^{-iz}$ when $z \gg 1$ along the positive real axis.  This fixes the parameter $\xi$ to
\beeq
\xi=\xi_++\xi_-,~~~\xi_+=c_{++}e^{i\pi j/2}-c_{+-},~~~\xi_-=c_{--} e^{-i\pi j/2}-c_{+-}~
%\label{}
\eneq 
where
\beeq
c_{\pm \pm}=\mathcal{C}  \int_0^\infty d\eta \Phi_\pm^{(0)} \delta V \Phi_\pm^{(0)}~.
%\label{}
\eneq 
In the large $z$ limit, we find 
\beeq
\Phi \sim -e^{-i\pi (1+j)/4} \sin(\pi j/2) \left(1-{\xi_- \over (-r_0\omega/c)^{4\over 5}} \right) e^{-iz} ~.
\label{largeZ-theta0}
\eneq

We now explore how things change when one wants to switch from the anti-Stokes line along the positive $z$ axis to the line along the negative $z$ axis.  We know that 
\beeq
\Phi_\pm^{(0)}\left(e^{i\theta}e^{-i\theta}z\right)= e^{i\theta (1\pm j)/2}\Phi_\pm^{(0)}\left(e^{-i\theta}z\right)~,
%\label{}
\eneq
where $\theta=\arg(z)$.  It is also easy to show that
\beeq
\delta V\left(e^{i\theta}e^{-i\theta}z\right)= e^{i\theta \left({4\over 5}-2\right)}\delta V\left(e^{-i\theta}z\right)~.
%\label{}
\eneq
(We are keeping the expansion power $4\over 5$ to be able to compare our results with those in \cite{M-S}.)  Using the two equations above, one can show
\beeq
\Phi_\pm^{(1)}\left(e^{i\theta}e^{-i\theta}z\right)= e^{i\theta \left({1 \over 2}+{4\over 5}\pm {j\over 2}\right)}\Phi_\pm^{(1)}\left(e^{-i\theta}z\right)~
%\label{}
\eneq
and
\begin{eqnarray}
\Phi\left(e^{i\theta}e^{-i\theta}z\right) &=& e^{i\theta (1+j)/2}   \left[ \Phi_+^{(0)}\left( e^{-i\theta}z\right) -e^{-i(\theta+\pi/2)j} \Phi_-^{(0)}\left( e^{-i\theta}z\right)  \right] \nonumber \\
 &&+  {e^{i\theta \left({1 \over 2}+{4\over 5}+ {j\over 2}\right)}\over (-r_0\omega/c)^{4\over 5}} \left[ \Phi_+^{(1)}\left( e^{-i\theta}z\right)  \right. \nonumber \\
&& \left. -e^{-i(\theta+\pi/2)j}\left(\Phi_-^{(1)}\left( e^{-i\theta}z\right) -\xi e^{-i \theta {4\over 5}} \Phi_-^{(0)} \left( e^{-i\theta}z\right) \right) \right] ~. 
\label{3pi-rotation}
\end{eqnarray}
We also have to determine how the change in $\theta$ in the $z$-plane corresponds to the change in angle in the $r$-plane.  When $r \rightarrow 0$, we have:
\beeq
z\approx -{\omega/c  \over 5r_0^4} r^{5}~,
%\label{}
\eneq
The anti-Stokes lines in the $z$-plane are the lines on which
\beeq
\pi+\arg{ \omega} + 5\arg{r}= l \pi~,
%\label{}
\eneq
where $l=0, 1, 2, \ldots$.  Note that we need to determine $\arg{r}$ on the two anti-Stokes lines that extend to infinity above and below the pole at the horizon.  These two lines are separated from each other by an angle of $\Delta \arg{r} = {3\pi /   5}$ as can be seen in Fig. \ref{3+1-BH}.  This translates to $\Delta \arg{z} = 3\pi $.  Therefore, if we choose our branch cuts in the $r$-plane and $z$-plane so that $\arg{z}=\theta=0$, we need to take our final $\theta$ to be $3\pi$.

For $\theta=0$ (positive real axis in $z$-plane), (\ref{3pi-rotation}) can be approximated in the limit $|z|(=e^{-i\theta}z) \rightarrow \infty$ with the expression given in Eq.\ (\ref{largeZ-theta0}).  
For $\theta=3\pi$ (negative real axis in $z$-plane), (\ref{3pi-rotation}) can be approximated in the limit $|z|(=e^{-i\theta}z) \rightarrow \infty$ with the expression
\begin{eqnarray}
\Phi (z) &\sim & e^{-i\pi (1+j)/4} \sin\left({3\pi j\over 2}\right) \left[1-{1 \over (-r_0\omega/c)^{4\over5}}A \right] e^{-iz}  \nonumber \\
&& + e^{i\pi (1-j)/4} \sin(2\pi j) \left[1-{1 \over (-r_0\omega/c)^{4\over 5}}B \right] e^{iz} ~,
\label{largeZ-theta3pi}
\end{eqnarray}
where
\begin{eqnarray}
A &=& {e^{-i\pi j/2}  \over 1-e^{3i\pi j}} \left[-c_{--}\left(1-e^{3i\pi {4\over5} }\right)-c_{++}e^{i\pi j}\left(1-e^{3i\pi \left({4\over 5} +j\right)}\right)\right. \nonumber \\
&& \left. +c_{+-} e^{i\pi j/2}\left(2-e^{3i\pi {4\over 5} }-e^{3i\pi \left({4\over 5} +j\right)}\right) \right] \nonumber \\
&=& {1-i \over 2}\left[ \xi_++i\xi_--\xi \cot(3\pi j/2) \right] \nonumber \\
&& + {1-ie^{2 i \pi/5} \over 2}\left[-i \xi_++i\xi_--\xi \cot(3\pi j/2)\right]~
%\label{}
\end{eqnarray}
and $B$ is a constant that contributes to the sub-dominant term and therefore is irrelevant in these calculations.
The two asymptotic solutions (\ref{largeZ-theta0}) and (\ref{largeZ-theta3pi}) at large $z$ or $r$ are related to one another via the monodromy at the event horizon, 
\begin{eqnarray}
\mathcal{M} &=&  - {\sin(3\pi j/2) \over \sin(\pi j/2)} \left\{1+   {1-i \over 2(-r_0\omega/c)^{4\over5}}\left[\xi_--\xi_++\xi \cot(3\pi j/2)\right] \right. \nonumber \\
&& \left. - {1-ie^{2 i \pi/5} \over 2(-r_0\omega/c)^{4\over5}}\left[-i \xi_++i\xi_--\xi \cot(3\pi j/2)\right]\right\}~.
%\label{}
\end{eqnarray}
Since the monodromy at the event horizon is\footnote{In \cite{Berti-sonic},the monodromy at the event horizon of a non-rotating canonical $3+1$-dimensional black hole is incorrectly taken to be $e^{4\pi  \omega }$ with $r_0=c=1$.  The correct monodromy for this case is $e^{\pi  \omega }$.  The monodromy in Eq.\ (\ref{monodromy}) reduces to $e^{\pi  \omega }$ when we take $r_0=c=1$. }
\beeq
\mathcal{M}= e^{2 i {\omega \over c} \oint {dr \over f(r)}}= e^{{\pi r_0 } \omega /c} 
\label{monodromy}
\eneq
where the contour of the integral is oriented clockwise around the pole at the horizon ($r=r_0$).  Combining the last two equations determines the QNM frequency spectrum in the large damping limit: 
\begin{eqnarray}
\pi r_0 \omega /c &=& \ln(1+2\cos(\pi j))+(2n+1)\pi i +  {1-i \over 2(-i)^{4\over5} (2n+1)^{4\over5}}\left[\xi_--\xi_++\xi \cot(3\pi j/2)\right] \nonumber \\
&& - {1-ie^{2 i \pi/5} \over 2(-i)^{4\over5} (2n+1)^{4\over5}}[i\xi_--i \xi_+-\xi \cot(3\pi j/2)]~.
%\label{}
\end{eqnarray}
In order to obtain an explicit expression for the QNM frequency spectrum above, we need the integral
\beeq
\mathcal{J}(\nu, \mu) \equiv \int_0^\infty dz z^{-1/5} J_\nu (z) J_\mu (z) = {\Gamma \left({1\over 5}\right) \Gamma \left({\nu +\mu +4/5 \over 2}\right) \over 2^{1/5} \Gamma \left({-\nu +\mu +6/5\over 2}\right) \Gamma \left({\nu +\mu+6/5 \over 2}\right)  \Gamma \left({\nu -\mu +6/5 \over 2}\right)}~
%\label{}
\eneq
which allows us to write
\beeq
c_{\pm \pm}=  \pi {25(1-j^2)+36 l(l+1) \over 72 \left(5^{6/ 5}\right) \sin(\pi j/2)}I(\pm j/2, \pm j/2)  ~.
%\label{}
\eneq
Using the above coefficients, we can determine $\xi_\pm$ and $\xi$ and finally, for $j=3/5$, we obtain the QNM frequency spectrum
\begin{eqnarray}
\pi r_0 \omega /c &=& \ln\left( {3\over 2}-{\sqrt{5} \over 2}\right)+(2n+1)\pi i  \nonumber \\
&& +  {2^{3/5}\left[ {1\over 5}(106+69 i)-(10-i)\sqrt{5} \right]^{1/5} (4+9l(l+1))\pi^2 \Gamma(1/5) \over 45 (2n+1)^{4/5} \Gamma(3/10)\Gamma^2(6/10)\Gamma(9/10)}~
%\label{}
\end{eqnarray}
at the large damping region.  Here, the leading order term is consistent with that obtained in \cite{Berti-sonic} but different by a factor of $4$.  This difference is explained in the footnote to Eq.\ (\ref{monodromy}).

\sxn{Quasinormal Modes in Bose-Einstein Condensate}

We follow the footsteps of Barcelo et al.\cite{Barcelo, Barcelo1} where they use the full Bogoliubov dispersion relation to perform stability analysis and calculate the QNMs for the $1+1$ dimensional (one dimensional flow) black hole in a BEC with step-like discontinuities at the sound horizon.  After substituting the Madelung representation in the Gross-Pitaevskii equation, one can linearize the equations.  This process leads to the following two sets of equations
\begin{eqnarray}
0&=&-\nabla \cdot (c^2 \bf{v}) ~,  \nonumber \\
0&=&-{1\over 2}m {\bf{v}}^2-mc^2-V_{\mbox{\scriptsize ext}}-\mu+{\hbar^2 \over 2m}{\nabla^2c \over c} ~
\label{continuity}
\end{eqnarray}
and
\begin{eqnarray}
\partial_t \tilde{n}_1&=&-\nabla \cdot (\tilde{n}_1{\bf{v}}+c^2 \nabla \theta_1)~, \nonumber \\
\partial_t \theta_1&=&-{\bf{v}}\cdot \nabla \theta_1-\tilde{n}_1+ {1\over 4} \xi^2 \nabla\cdot \left[c^2 \nabla \left({\tilde{n}_1 \over c^2}\right)\right] ~
\label{main-eqns}
\end{eqnarray}
where $c$ is the speed of sound, $\bf v$ is the flow velocity, $m$ is the boson mass, $V_ {\mbox{\scriptsize ext}}$ is the external potential, $\mu$ is the chemical potential, $\tilde{n}_1$ and $\theta_1$ are small perturbations of the density and phase of the BEC and $\xi \equiv {\hbar \over mc}$ is the so-called healing length.
Let us assume that the time dependency of the perturbations is of the form
\begin{eqnarray}
\tilde{n}_1 ({\bf{r}},t)&=& e^{-i\omega t}N_1({\bf{r}})~ \nonumber \\
\theta_1({\bf{r}},t)&=& e^{-i\omega t}\Theta_1({\bf{r}})~
\label{time-depend}
\end{eqnarray}
with constant $\omega$.
This gives
\begin{eqnarray}
-i\omega N_1&=&-\nabla \cdot (N_1{\bf{v}}+c^2 \nabla \Theta_1)~,  \\
-i \omega \Theta_1&=&-{\bf{v}}\cdot \nabla \Theta_1-N_1+ {1\over 4} \xi^2 \nabla\cdot \left[c^2 \nabla \left({N_1 \over c^2}\right)\right] ~.
\label{eq-quantum-term}
\end{eqnarray}
We are interested in the large $\omega$ limit.  In this limit, the only term that can be ignored in the above equations is the second term ($-N_1$) in the right hand side of Eq.\ (\ref{eq-quantum-term}).  No further simplification can be done.  However, if we seek planewave solutions
\begin{eqnarray}
N_1({\bf{r}})&=& A e^{i \bf{k\cdot r}}~ \nonumber \\
\Theta_1({\bf{r}})&=& Be^{i \bf{k\cdot r}}~
%\label{}
\end{eqnarray} 
in a homogeneous region of the BEC where $c$ and $v$ are constants, we get two sets of equations that can be written in matrix form
\beeq
\left( \begin{array}{ll}
                   i(\omega-{\bf{k\cdot v}})~~~~~~~c^2k^2\\
                   1+{1\over 4} \xi^2k^2~~~-i(\omega-{\bf{v\cdot k}})
                   \end{array}
           \right)  
\left( \begin{array}{ll}
                   A\\
                   B
                   \end{array}
           \right)		=0	~.				
\label{WKB}
\eneq
To have a non-trivial solution to this set of equations, the determinant of the above matrix must vanish.  This leads to the full Bogoliubov dispersion relation:
\beeq
(\omega-{\bf{v\cdot k}})^2=c^2k^2+{1\over 4}c^2 \xi^2 k^4~.
%\label{}
\eneq
This dispersion relation is identical to that obtained in the hydrodynamic limit but with the addition of a quantum term involving the healing length $\xi$.
Solving for $\omega$ leads to
\beeq
\omega={\bf{v\cdot k}}\pm \sqrt{c^2k^2+{1\over 4}c^2 \xi^2 k^4}~.
%\label{}
\eneq
It is clear that in the limit where $|\omega| \rightarrow \infty$, we have to require $|k| \rightarrow \infty$.  In this case 
\beeq
\omega \approx \pm {1\over 2}c \xi k^2=\pm {\hbar \over 2m} k^2~.
%\label{}
\eneq
Note that the only term that survives in this approximation is the quantum term that involves the healing length $\xi$, which is the quantum length scale of the system.  We see that, similar to gravitational black holes, the small scale structure of the background medium in which analog black holes are formed can also be probed using high overtone QNMs.   

The new simplified equations when $|\omega | \rightarrow \infty $ are
\begin{eqnarray}
-i\omega N_1 &\approx & -\nabla \cdot (c^2 \nabla \Theta_1)~,  \label{eqn: frobenius 6}\\
-i \omega \Theta_1  &\approx &   {1\over 4} \xi^2 \nabla\cdot \left[c^2 \nabla \left({N_1 \over c^2}\right)\right] ~. 
\label{eqn: frobenius 7}
\end{eqnarray}
%These two equations can further be combined into one linear differential equation of fourth order. 

We now will follow the method that was developed in \cite{Barcelo, Barcelo1} to see if there exist high overtone QNMs in the one dimensional case represented in FIG. 1 of \cite{Barcelo1}.   Integrating the above equations in an infinitesimal neighborhood of the discontinuity (taken to be at $x=0$) and simplifying with the use of the continuity equation $vc^2=$constant [see Eq.\ (\ref{continuity})] leads to the following sets of matching conditions:
\begin{eqnarray}	
\left[c^2\partial_x \Theta_1\right]&=& 0 \\
\left[\Theta_1\right]&=&0  \\
\left[c^2 \partial_x \left({N_1 \over c^2}\right)\right]&=&0 \\
\left[{N_1 \over c^2}\right]&=&0~. 
%\label{}
\end{eqnarray}
where the brackets indicate, for example, $[\Theta_1]=\Theta_1|_{x=0^+}-\Theta_1|_{x=0^-}$.  The first and third equations above are found from integrating (\ref{eqn: frobenius 6}) and (\ref{eqn: frobenius 7}) and the others are the result of a second integration.  Note that these conditions, which are based on the high overtone quasinormal frequency limit, are exactly the same as the conditions we would obtain from integrating the exact equations in (\ref{main-eqns}). 

We now assume planewave solutions where: 
\beeq
N_1=
\left\{\begin{array}{ll}
			 \sum_{j=1}^4 A_j e^{i k_j x}~~~(x<0)~, \\
       \sum_{j=5}^8 A_j e^{i k_j x}~~~(x>0)~, 
       \end{array} \right.	
%\label{}
\eneq

inserting the above solution into the wave equations gives
\beeq
\Theta_1\approx
\left\{\begin{array}{ll}
			 \sum^{4}_{j=1}{{\omega \over ic_L^2 k_j^2}e^{i k_j x}}~~~(x<0)~, \\
       \sum_{j=5}^8 A_j {\omega \over ic_R^2 k_j^2}e^{i k_j x}~~~(x>0)~, 
       \end{array} \right.
%\label{}
\eneq
\beeq
\Theta_1\approx \sum_j A_j {\omega \over ic^2 k_j^2}e^{i k_j x} ~,
%\label{}
\eneq
Here we choose $k_1=\sqrt{2m\omega / \hbar}$, $k_2=-\sqrt{2m\omega / \hbar}$, $k_3=i\sqrt{2m\omega / \hbar}$ and $k_4=-i\sqrt{2m\omega / \hbar}$ for the region $x<0$ and $k_5=\sqrt{2m\omega / \hbar}$, $k_6=-\sqrt{2m\omega / \hbar}$, $k_7=i\sqrt{2m\omega / \hbar}$ and $k_8=-i\sqrt{2m\omega / \hbar}$ for the region $x>0$.

\beeq
v_g= \mbox{Re} \left({d\omega \over dk} \right)=\mbox{Re} \left( {c^2k+{1\over 2}\xi^2c^2k^3 \over \omega -vk}+v\right)\rightarrow  \mbox{Re} \left( {{1\over 2}\xi^2c^2k^3 \over \omega }\right)~,
%\label{}
\eneq
The real part of $\omega$ needs to be a positive value and, considering our choice of time dependency in Eq.\ (\ref{time-depend}), the imaginary part needs to be negative. It is now easy to show that the outgoing waves at $-\infty$ are associated with $k_2$ and $k_4$ and at $+\infty$ they are associated with $k_5$ and $k_7$.  Applying the boundary conditions at $\pm\infty$ will give us four more conditions ($A_1=A_3=A_6=A_8=0$) in addition to the four conditions at the discontinuity.  These conditions can be written in matrix form as below:
\beeq
\left( \begin{array}{ll}
                   ~~{\omega \over k_1}~~~~~{\omega \over k_2}~~~~~{\omega \over k_3}~~~~~{\omega \over k_4}~~~~~~~-{\omega \over k_5}~~~~~-{\omega \over k_6}~~~~~-{\omega \over k_7}~~~~~-{\omega \over k_8}\\
                   {\omega \over c_L^2k_1^2}~~~{\omega \over c_L^2k_2^2}~~~{\omega \over c_L^2k_3^2}~~~{\omega \over c_L^2k_4^2}~~~-{\omega \over c_R^2k_5^2}~~~-{\omega \over c_R^2k_6^2}~~~-{\omega \over c_R^2k_7^2}~~~-{\omega \over c_R^2k_8^2}\\
									 ~~{k_1}~~~~~{k_2}~~~~~{ k_3}~~~~~{ k_4}~~~~~~~-{ k_5}~~~~~-{k_6}~~~~~-{ k_7}~~~~~-{ k_8}\\
									~~{1 \over c_L^2}~~~~~{1 \over c_L^2}~~~~~{1\over c_L^2}~~~~~{1 \over c_L^2}~~~~~~-{1 \over c_R^2}~~~~~-{1 \over c_R^2}~~~~~-{1 \over c_R^2}~~~~~-{1 \over c_R^2}\\
									~~{1 }~~~~~~~{0}~~~~~~~{0}~~~~~~{0}~~~~~~~~~~~{0}~~~~~~~~~{0}~~~~~~~~~~{0}~~~~~~~~~~{0}\\
									~~{0}~~~~~~~{0}~~~~~~~{1}~~~~~~{0}~~~~~~~~~~~{0}~~~~~~~~~{0}~~~~~~~~~~{0}~~~~~~~~~~{0}\\
									~~{0}~~~~~~~{0}~~~~~~~{0}~~~~~~{0}~~~~~~~~~~~{0}~~~~~~~~~{1}~~~~~~~~~~{0}~~~~~~~~~~{0}\\
									~~{0}~~~~~~~{0}~~~~~~~{0}~~~~~~{0}~~~~~~~~~~~{0}~~~~~~~~~{0}~~~~~~~~~~{0}~~~~~~~~~~{1}\\
                   \end{array}
           \right)  
\left( \begin{array}{ll}
                   A_1\\
                   A_2\\
									 A_3\\A_4\\A_5\\A_6\\A_7\\A_8\\
                   \end{array}
           \right)		=0					
%\label{WKB}
\eneq
The determinant of this matrix should be equal to zero for non-trivial solutions.  This gives:
\beeq
\left( { 1 \over c_L^4}+{1 \over c_R^4} \right) 2 i \hbar \omega =0 ~,
%\label{}
\eneq
This clearly gives us $\omega = 0$, which means the QNMs do not exist in this particular one-dimensional black hole configuration.  This is not the only case where high overtone QNMs do not exist.  It was shown in \cite{Berti-sonic}  that QNMs do not exist in the rotating $2+1$-dimensional acoustic black holes as well.  We have also searched numerically in the high overtone area and we found no QNMs.  In other words, our numerical result is consistent with the analytical result above.

\sxn{Summary and Conclusion}

We calculated the leading order and the first order correction of the high overtone QNMs of a $3+1$ dimensional canonical non-rotating acoustic black hole.  We also showed that the high overtone QNMs do not exist in the $1+1$ dimensional black hole in the background of a BEC.  It is, however, clear that in the high overtone limit the quantum potential, which is a small scale effect, dominates over all the other terms.  In the case of BEC, the quantum potential significantly changes the QNM spectrum in the case of a $1+1$ dimensional black hole as was pointed out by the authors of \cite{Barcelo, Barcelo1}.  Therefore, one does not need to go to the high overtone limit to study such a potential. But by looking at the high overtone part of the QNM spectrum, one can make sure that all the observed effects are due to the microscopic structure of the BEC rather than other macroscopic effects.  It is also possible to detect other unknown small scale effects using high overtone QNMs. A better understanding of the small scale structure of quantum fluids such as BEC will lead to a better understanding of the macroscopic behavior of these many-body quantum systems in terms of their individual components. 

%For example, it is possible that in a quantum fluid like BEC the Generalized Uncertainty Principle (see for example \cite{Maggiore-GUP} for one of the early papers on this topic) may lead to a correction to the Gross-Pitaevskii equation.  This correction could potentially be detected using high overtone QNMs.
%At this point, we would like to suggest an speculative idea regarding a possible example for an interesting unknown small scale effect, which may be explored using high overtone QNMs.  We speculate that an intriguing microscopic correction to the Gross-Pitaevskii equation may be due to the Generalized Uncertainty Principle (see for example \cite{Maggiore-GUP} for one of the early papers on this topic), which is a quantum gravity correction to Heisenberg's Uncertainty Principle.  
%To put this matter on a firmer ground, such corrections to the Gross-Pitaevskii equation need to be derived and the high overtone QNMs with such corrections need to be explored to see if they can lead to an experimentl observation of such quantum gravity effects.     
Future steps are to explore the high overtone QNMs of $2+1$ and $3+1$ dimensional black hole configurations formed in a BEC.  We believe such modes should exist in the $3+1$ dimensional configuration since they exist in the hydrodynamic limit\cite{Berti-sonic}.

%\newpage

% A useful Journal macro
\def\jnl#1#2#3#4{{#1}{\bf #2} (#3) #4}

\def\Zphys{{\em Z.\ Phys.} }
\def\jssc{{\em J.\ Solid State Chem.\ }}
\def\jpsJ{{\em J.\ Phys.\ Soc.\ Japan }}
\def\ptps{{\em Prog.\ Theoret.\ Phys.\ Suppl.\ }}
\def\PTP{{\em Prog.\ Theoret.\ Phys.\  }}
\def\LNC{{\em Lett.\ Nuovo.\ Cim.\  }}

\def\JMP{{\em J. Math.\ Phys.} }
\def\NPB{{\em Nucl.\ Phys.} B}
\def\NP{{\em Nucl.\ Phys.} }
\def\PLB{{\em Phys.\ Lett.} B}
\def\PL{{\em Phys.\ Lett.} }
\def\PRL{\em Phys.\ Rev.\ Lett. }
\def\PRA{{\em Phys.\ Rev.} A}
\def\PRB{{\em Phys.\ Rev.} B}
\def\PRD{{\em Phys.\ Rev.} D}
\def\PR{{\em Phys.\ Rev.} }
\def\PRe{{\em Phys.\ Rep.} }
\def\PMP{{\em Philos.\ Mod.\ Phys.} }
\def\AP{{\em Ann.\ Phys.\ (N.Y.)} }
\def\RMP{{\em Rev.\ Mod.\ Phys.} }
\def\ZPC{{\em Z.\ Phys.} C}
\def\SCI{\em Science}
\def\CMP{\em Comm.\ Math.\ Phys. }
\def\MPLA{{\em Mod.\ Phys.\ Lett.} A}
\def\IJMPB{{\em Int.\ J.\ Mod.\ Phys.} B}
\def\IJMPA{{\em Int.\ J.\ Mod.\ Phys.} A}
\def\cmp{{\em Com.\ Math.\ Phys.}}
\def\JPA{{\em J.\  Phys.} A}
\def\CQG{\em Class.\ Quant.\ Grav.~}
\def\ATMP{\em Adv.\ Theoret.\ Math.\ Phys.~}
\def\PRSA{{\em Proc.\ Roy.\ Soc.\ Lon.} A }
\def\IJTP{\em Int.\ J.\ Theor.\ Phys.~}
\def\ibid{{\em ibid.} }
\def\LRR{{\em Living \ Rev.\ Relative.} }
\def\FP{{\em Found.\ Phys.} }
\vskip 1cm

%\leftline{\bf References}

\renewenvironment{thebibliography}[1]
        {\begin{list}{[$\,$\arabic{enumi}$\,$]}  
% {\arabic{enumi}.}
        {\usecounter{enumi}\setlength{\parsep}{0pt}
         \setlength{\itemsep}{0pt}  \renewcommand{\baselinestretch}{1.2}
         \settowidth
        {\labelwidth}{#1 ~ ~}\sloppy}}{\end{list}}

\end{document}